\documentclass{ieeeaccess}
\usepackage{cite}
\usepackage{amsmath,amssymb,amsfonts,physics}
\usepackage{cleveref}
\usepackage{algorithmic}
\usepackage{graphicx}
\usepackage{textcomp}
\usepackage{array}

\usepackage{bm}
\usepackage{gensymb}

\usepackage{soul}
\usepackage{xcolor}
\definecolor{warmpastelpurple}{RGB}{235, 210, 255}
\sethlcolor{warmpastelpurple}

\makeatletter
\AtBeginDocument{\DeclareMathVersion{bold}
	\SetSymbolFont{operators}{bold}{T1}{times}{b}{n}
	\SetSymbolFont{NewLetters}{bold}{T1}{times}{b}{it}
	\SetMathAlphabet{\mathrm}{bold}{T1}{times}{b}{n}
	\SetMathAlphabet{\mathit}{bold}{T1}{times}{b}{it}
	\SetMathAlphabet{\mathbf}{bold}{T1}{times}{b}{n}
	\SetMathAlphabet{\mathtt}{bold}{OT1}{pcr}{b}{n}
	\SetSymbolFont{symbols}{bold}{OMS}{cmsy}{b}{n}
	\renewcommand\boldmath{\@nomath\boldmath\mathversion{bold}}}
\makeatother

\def\BibTeX{{\rm B\kern-.05em{\sc i\kern-.025em b}\kern-.08em
		T\kern-.1667em\lower.7ex\hbox{E}\kern-.125emX}}

\begin{document}
	\history{Date of publication xxxx 00, 0000, date of current version xxxx 00, 0000.}
	\doi{10.1109/ACCESS.2024.0429000}
	
	\title{Differential Evolution-Based End-Fire Realized Gain Optimization of Active and Parasitic Arrays}
	\author{\uppercase{Rozita Konstantinou}\authorrefmark{1}, 
		\uppercase{Ihsan Kanbaz}\authorrefmark{1,2}, \uppercase{Okan Yurduseven}\authorrefmark{1} \IEEEmembership{Senior Member, IEEE}, and \uppercase{Michail Matthaiou}\authorrefmark{1} \IEEEmembership{Fellow, IEEE}}
	
	\address[1]{Centre for Wireless Innovation (CWI), Queen’s University Belfast, Belfast, U.K.}
	\address[2]{Electrical and Electronics Engineering, Gazi University, Ankara, Turkey}
	\tfootnote{This work was supported by the European Research Council (ERC) under the European Union’s Horizon 2020 research and innovation programme (grant agreement No. 101001331). The work of O. Yurduseven was supported by a research grant from the Leverhulme Trust under the Research Leadership Award RL-2019-019.}
	
	\markboth
	{Konstantinou \headeretal: Preparation of Papers for IEEE TRANSACTIONS and JOURNALS}
	{Konstantinou \headeretal: Preparation of Papers for IEEE TRANSACTIONS and JOURNALS}
	
	\corresp{Corresponding author: Rozita Konstantinou (e-mail: rkonstantinou01@qub.ac.uk).}

	\begin{abstract} 
		We propose a novel approach for boosting the realized gain of arrays with enhanced directivity, utilizing both active and parasitic dipoles. The optimization process first maximizes the end-fire gain in the active array by selecting the optimal current excitation vector. For the parasitic arrays, the dipoles are reactively loaded based on the input impedances of the active dipoles, after which the optimization focuses on the inter-element distance to achieve a balance between the gain and the reflection efficiency. This multi-objective optimization, underpinned by the differential evolution (DE) algorithm, uses a simple wire dipole as the unit element. Full-wave simulations validate our theoretical results, showing that our two- and three-element parasitic arrays achieve realized gain comparable to state-of-the-art designs without relying on intricate unit elements or resource-intensive simulations, while our four- and five-element parasitic arrays yield the highest realized gain values reported in the literature. The simplicity of our approach allows optimizations to run significantly faster than full-wave simulations, whilst the sensitivity analysis showcases the robustness of the design under small deviations in loads and element positioning. Compact and power-efficient, the proposed parasitic arrays are well-suited for base stations, aligning with modern communication system requirements while minimizing hardware complexity.
	\end{abstract}
	
	\begin{keywords}
		Active antenna arrays, differential evolution, end-fire arrays, high efficiency, high realized gain, parasitic antenna arrays.
	\end{keywords}
	
	\titlepgskip=-21pt
	
	\maketitle
	
	\section{Introduction}
	\label{sec:introduction}
	\PARstart{A}{ntenna} arrays are typically designed with uniform element spacing, which is at least a half wavelength, to mitigate mutual coupling, which kicks in when the power fed to one element of the array is induced to the other elements. However, when the element spacing exceeds half a wavelength, grating lobes emerge in the radiation pattern, which can degrade the performance of the array by directing energy in unwanted directions \cite{globes}. When maximum ratio transmission is implemented, a conventional phased array with $N$ isotropic elements yields a gain that is linearly proportional to $N$ \cite{ref10}. On the other hand, exploring reduced element spacing and leveraging mutual coupling introduces the concept of super-directivity, offering the potential for higher array gains \cite{marz1,KD}. In particular, it has been theoretically established that the directivity of an array of isotropic elements increases super-linearly with the number $N$ of antennas as the spacing between the elements decreases \cite{bell}. Theoretical studies from as early as 1946 demonstrated that as the distance between $N$ isotropic radiating elements tends to zero, the maximum end-fire directivity of the linear array could theoretically scale up as $N^2$ \cite{uzkov}. Some recent research results, using spherical wave expansion (SWE), extended this limit for electrical dipoles to $N^2 + N - 1/2$ \cite{delav1}.
	
	While decreasing inter-element distance can enhance directivity, it may hinder the antenna’s performance by adversely impacting the antenna matching and radiation efficiency \cite{kalis}. Placing the antennas closer together can lead to impedance mismatch due to complex factors, such as surface currents and near-field radiations, which are difficult to calculate accurately \cite{nossek}. Similarly, not taking impedance mismatch into consideration during the design stages, can be detrimental to the reflection efficiency. In principle, choosing the optimal antenna excitation currents in conjunction with the right inter-element spacing can mitigate these issues by amplifying the radiated power, thus achieving super-gain and high efficiency \cite{nossek2}. In this work, we address these issues for active arrays by optimizing the antenna excitation currents and inter-element spacing to achieve high efficiency and enhanced realized gain.
	
	The deployment of parasitic sub-arrays is a reasonable approach to achieve satisfactory directivity compared to fully driven arrays while saving energy by utilizing significantly fewer digital-to-analog converters (DACs), radio frequency (RF) chains, and power amplifiers (PAs) \cite{RFs}. Parasitic arrays can maintain high performance without requiring intricate matching networks, as only the active element needs to be matched. Additional benefits include reduced complexity, as feeding networks are no longer required, and antenna size reduction.
	
	To date, the focus of most parasitic antenna array optimizations appears to have centered around enhancing the directivity or gain without incorporating the reflection efficiency into the process \cite{delav1,delav2,umma,DE0,DE00,DE1,delav3}. However, a low reflection (mismatch) efficiency is detrimental to the realized gain, which is the gain achievable in real-world conditions. This paper addresses that issue by focusing on optimizing both the directivity and the reflection efficiency to maximize the realized gain in parasitic arrays. In this space, in \cite{DE3}, the multi-objective function of the optimization for an S-monopole array considered both the directivity and input reflection coefficient. The authors employed the network characteristic modes (NCM) method, which relies on full-wave simulations, to obtain the impedance matrix and the electric field of each element. These parameters are used as inputs for the optimization process. Moreover, in \cite{delav4}, a SWE approach was employed, wherein the electromagnetic simulations were used to calculate the active pattern of each element within the array to allow for the determination of the optimal weighting coefficients for maximizing gain. In addition, in \cite{ziol2}, the unit element was optimized to achieve exceptionally high realized gain. The final array incorporates varying dimensions between the two electric dipoles and introduces a magnetic dipole, further enhancing the realized gain. 
	Unlike previous methods, our approach relies on the analytically calculated impedance matrices and the corresponding excitation vectors as the input of the optimization algorithm, while the full-wave solver is used solely to verify the efficacy of the proposed method. This is significant because it greatly reduces the computational burden and expedites the design process, making our method more efficient and accessible compared to those that depend on full-wave simulations for every iteration. The unit element used is a simple half-wavelength wire dipole, replicated to form an array configuration.
	Furthermore, our research addresses a gap in the literature concerning array-level mutual coupling-aware optimization, focusing explicitly on enhancing the realized gain of the array without the need for complex unit element optimizations. 
	
	In the realm of antennas and RF communications, multi-objective optimization algorithms have become instrumental. For instance, genetic algorithms (GAs) were used to maximize the realized gain of a circularly polarised end-fire electrically small antenna array in \cite{GAcirc}, and to determine the loading values of the parasitic elements employed in an electronically steerable passive array radiator (ESPAR) antenna, in order to realize a set of predefined radiation patterns, in \cite{GAespar}. Additionally, the authors of \cite{PSOpyr} deployed a particle swam optimization (PSO) algorithm to minimize back lobes in a uniform planar antenna array. Notably, in \cite{ihsan}, a differential evolution (DE) process was used to optimize the length, spacing, and excitation of a wire dipole array, achieving super-realized gain. Note that DE is widely applied as a method across numerous studies. In particular, the researchers in \cite{DE0} introduced a parasitic dipole antenna optimized to achieve super-directivity and high radiation efficiency. It is noteworthy that all the studied topologies referenced above have been limited to either active or parasitic configurations. Moreover, \cite{DE00,DE1,DE3} utilized the NCM approach to construct both active and parasitic super-directive arrays with an S-shaped monopole unit element. In the presented work, we conceive an optimization framework based on DE and tailored for both active and parasitic dipole arrays, with the emphasis on enhancing the realized gain and consequently, the overall efficiency. 
	
	The key contributions of this paper are outlined below:
	\begin{itemize}	
		\item We incorporate the total efficiency into the optimization process, with the realized gain as the primary focus. We introduce an active antenna array model that accounts for mutual coupling and losses, calculating the realized gain based on the combined reflection coefficients of each dipole element. We then present the necessary modifications to this model to study a parasitic array with one fed element. This approach ensures that the proposed model and optimizations result in efficient antenna designs.
		\item We propose a versatile optimization framework applicable to both active and parasitic array architectures. Our approach uses a DE algorithm to optimize the positioning and feed excitation of the active arrays, as well as the spacing and loads for the parasitic arrays, providing a unified approach for antenna array design.
		\item Our approach notably reduces the computational burden by utilizing a theoretical framework as the input for the optimization process, rather than relying on full-wave simulations, making the design process more efficient. Additionally, we provide a detailed runtime comparison to quantify the significant time savings achieved by this method.
		\item We perform a sensitivity analysis on the five-element parasitic array, examining variations in both loads and element positions. This analysis confirms the robustness of the proposed optimization framework, ensuring that the optimized designs sustain their satisfactory performance even under realistic parameter deviations.
		\item We illustrate the efficacy of the proposed optimization approach, highlighting the advantages and trade-offs of the optimized active and parasitic arrays compared to a conventional uniform linear array (ULA). The design parameters obtained through our method are then validated with full-wave electromagnetic simulations, which confirm the accuracy of our theoretical analysis. Finally, we compare the arrays produced by our method with those presented in the literature, highlighting the advantages and effectiveness of our approach.
	\end{itemize}
	
	The paper is organized as follows: In \Cref{sec:model}, the array model is outlined, for both the active and the parasitic case. \Cref{sec:optimization} introduces the optimization process that is based on DE and presents the numerical results. \Cref{sec:simulations} is devoted to the full-wave simulations and the interpretation of the results along with the theoretical and simulated runtime comparisons and sensitivity analysis. In \Cref{sec:discussion}, the optimized parasitic arrays are compared to the state of the art. Finally, \Cref{sec:conclusion} outlines the primary conclusions.
	
	\textit{Notations}: Throughout the paper, ${\bf A}(\cdot,\cdot,\cdot)$ is a vector field; ${\bf A}$ is a matrix; ${\bf A}^T$ and ${\bf A}^H$ are the transpose and conjugate transpose of ${\bf A}$, respectively; ${\bf A}_{i,j}$ is the $(i,j)^{th}$ entry of ${\bf A}$; ${\bf a}$ is a vector; ${\hat{\bf a}}$ is the unit vector of ${\bf{a}}$; $\|{\bf{a}}\|$ is the $l_2$-norm of {\bf{a}}; ${\bf{I}}_N$ is the $N \times N$ identity matrix; ${\bf{a}} \cdot {\bf{b}}$ is the inner product between ${\bf{a}}$ and ${\bf{b}}$. Finally, $\Re\{\cdot\}$ and $\Im\{\cdot\}$ are the real and imaginary parts of a complex variable, respectively.
	
	\section{Theoretical Background}\label{sec:model}
	\subsection{Active Antenna Array Model}\label{sec:actmod}
	\begin{figure}[htbp]
		\centering{\includegraphics[scale=0.6]{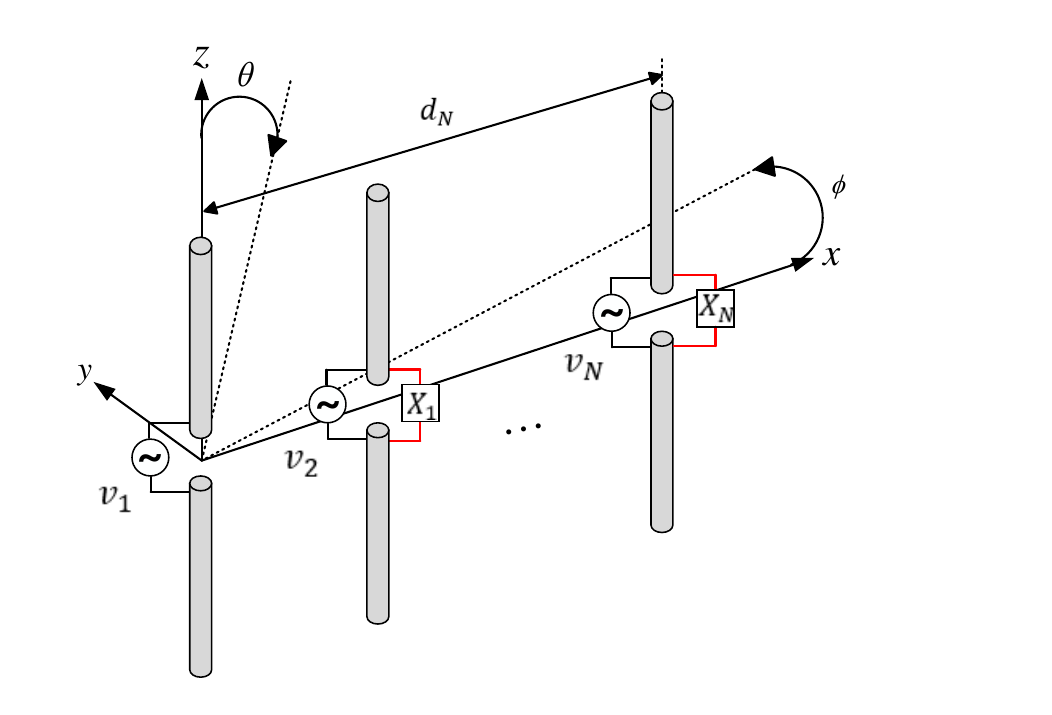}}
		\caption{Geometry of the combined active and parasitic antenna array configurations. The active array shows the elements connected to the excitation source (black lines) for elements $n>1$. For the parasitic array, these elements are connected to their respective loads through red lines.}\label{dipoles}
	\end{figure}
	
	To begin with, the model of the antenna array consisting of $N$ fed and coupled antennas parallel to the $z$-axis, with a length of $l$, and placed along the $x$-axis (see Fig.~\ref{dipoles}), is presented. For this architecture, the spherical coordinates $(r, \theta, \phi)$ are used, where $r$ is the distance to the point from the origin, $\theta \in [0,\pi]$ is the elevation angle and $\phi \in [-\pi,\pi]$ is the azimuth angle, of a point $(x, y, z)$ in the Cartesian coordinate system. The unit vector of $r$ is ${\hat{\bf r}} = \sin\theta \cos\phi \ {\hat{\bf x}} + \sin\theta \sin\phi  \ {\hat{\bf y}} + \cos\theta \ {\hat{\bf z}}$.
	
	\begin{table}[!h]
		\renewcommand{\arraystretch}{1.3}
		\caption{Model Simulation Parameters}
		\label{tablemodel}
		\centering
		\begin{tabular}{| c | c | c |}
			\hline
			Notation & Parameter & Value \\
			\hline
			$f$ & carrier frequency & $3.5\ GHz$ \\
			$l$ & dipole length & $\lambda/2$ \\
			$\rho$ & dipole radius & $\lambda/200$ \\
			$k$ & wave number & $2\pi/\lambda$ \\
			$\eta$ & impedance of free-space & $\approx120\pi$ \\
			$\mu$ & permeability of free-space & $4\pi \times 10^{-7} \ H/m$ \\
			$\sigma_c$ & conductivity of copper & $5.8\times10^{7} \ S/m$ \\
			$P_t$ & maximum input power & $0.5 \ W$ \\
			\hline
		\end{tabular}
	\end{table}
	
	For the studied array topology depicted in Fig.~\ref{dipoles}, the system parameters given in Table~\ref{tablemodel} are used. The operating frequency is selected to be $3.5$ GHz \cite{freq1,freq2}, which is within the sub-$6$ GHz $5$G spectrum in the UK. We also note that the presented optimization framework in this paper can be applied to other frequency bands without loss of generality. The complex input currents and voltages of the coupled dipoles, assuming they are lossless, are connected via the following relationship:
	\begin{equation}
		{\bf v}' = {\bf Z}' \cdot {\bf i} \label{v1},
	\end{equation}
	where ${\bf Z}' \in \mathbb{C} ^{N \times N}$ is the symmetric mutual impedance matrix with $Z_{11}'=\hdots=Z_{NN}'$, calculated as in \cite{orf}, ${\bf i} = \begin{bmatrix} i_1,\hdots, i_N \end{bmatrix}^T \in \mathbb{C} ^{N \times 1}$ and ${\bf v}' = \begin{bmatrix} v_1',\hdots, v_N' \end{bmatrix}^T \in \mathbb{C} ^{N \times 1}$.
	The array consists of very thin finite-length linear dipole elements.  The total radiated power of the array is as follows \cite{orf}:
	\begin{equation}
		P_{rad} =\frac{1}{2} {\bf{i}}^H \cdot \Re\{{\bf{Z}}'\} \cdot {\bf{i}}. \label{Prad}
	\end{equation} 
	The far-field electric field of the array at $(r, \theta, \phi)$ is described by \cite{balanis}:
	\begin{equation}
		{\bf{E}}(r,\theta,\phi) = j\eta \frac{e^{-jkr}}{2\pi r} {\bf{F}}(\theta) \displaystyle\sum_{n=1} ^{N} e^{jk{\hat{\bf r}}\cdot {\bf{r}}_{n}} \cdot i_{n},
	\end{equation}
	where
	\begin{equation}		
		{\bf{F}}(\theta) = \frac{\cos\left( \frac{kl}{2} \cos\theta \right) - \cos\left( \frac{kl}{2} \right)}{\sin\theta} {\hat{\bf\theta}} \label{F},
	\end{equation}
	is the element factor of each dipole, while ${\bf{r}}_{n}=(d_n,0,0)$ is the position vector of the $n^{th}$ element. In addition, ${\bf{F}}(\theta)$ represents the isolated element pattern (IEP) and is used in place of the active element pattern (AEP). The IEP represents the radiation pattern of a single element in isolation, while the AEP reflects the radiation pattern of the same element, influenced by nearby elements due to mutual coupling \cite{aep}. While implementing the IEP in the model may be less accurate than utilizing the AEP, it still yields satisfactory results without the intricacies involved in integrating the AEP, which necessitates numerous full-wave electromagnetic simulations, a process that may require a significant amount of computational time.
	
	One factor influencing the efficiency and performance of realistic antennas is the conduction/loss resistance, which results in the dissipation of heat. Given the sinusoidal current distribution of the element, the loss resistance relative to the input current $R_{loss}$ is computed as in \cite{KD}. The overall power loss is:
	\begin{equation}	
		P_{loss} = \frac{1}{2} R_{loss} \norm{{\bf i}}^2 = \frac{1}{2} {\bf{i}}^H \cdot {\bf R}_{loss} \cdot {\bf i},
	\end{equation}
	where ${\bf R}_{loss} = R_{loss} {\bf{I}}_N \in \mathbb{R} ^{N \times N}$. The input power at the antenna ports is the sum of the power loss and the radiated power \cite{bird}:
	\begin{equation}
		\begin{aligned}
			P_{in} = \frac{1}{2} {\bf{i}}^H \cdot {\bf R}_{loss} \cdot {\bf i} + \frac{1}{2} {\bf{i}}^H \cdot \Re\{{\bf{Z}}'\} \cdot {\bf{i}} = \frac{1}{2} {\bf{i}}^H \cdot \Re\{{\bf{Z}}\} \cdot {\bf{i}} \label{Pin},
		\end{aligned}
	\end{equation}
	where
	\begin{equation}
		\begin{aligned}	
			{\bf{Z}} = {\bf{Z}}' + {\bf R}_{loss} \in \mathbb{C} ^{N \times N} \label{Zmatrix},
		\end{aligned}
	\end{equation} 
	is the impedance matrix of the lossy array. The input voltage vector is now:
	\begin{equation}
		{\bf v} = {\bf Z} \cdot {\bf i} \label{v2}.
	\end{equation}
	
	Mutual coupling affects the input impedance of each element, making the active impedance of each dipole \cite{balanis}:
	\begin{equation}	
		\begin{aligned}
			Z_{in,n} = Z_{act,n} = \frac{v_n}{i_n} = Z_{nn} + \displaystyle\sum_{m=1,m \neq n} ^{N} Z_{nm}\frac{i_m}{i_n} \label{Zact}.	
		\end{aligned}			
	\end{equation}
	
	A few different efficiencies are linked to each antenna. The total antenna efficiency accounts for the conduction and dielectric losses, on top of the reflection losses due to the mismatch between the transmission lines and the ports of the antenna. It can be written as $e_0=e_re_ce_d$, where $e_r$, $e_c$ and $e_d$ are the reflection, conduction, and dielectric efficiency, respectively. The conduction and dielectric efficiencies are commonly combined and comprise the radiation efficiency, $e_{cd}=e_ce_d$, which establishes the connection between the gain and directivity and is computed as:
	\begin{equation}
		\begin{aligned}	
			e_{cd} = \frac{P_{rad}}{P_{in}} = \frac{G(\theta,\phi)}{D(\theta,\phi)} = \frac{{\bf{i}}^H \cdot \Re\{{\bf{Z}}'\} \cdot {\bf{i}}}{{\bf{i}}^H \cdot \Re\{{\bf{Z}}\} \cdot {\bf{i}}} \label{ecd1},
		\end{aligned}
	\end{equation}
	where $D(\theta,\phi)$ is the directivity of the array and
	\begin{align}	
		G(\theta,\phi) = \frac{\eta}{\pi}\| {\bf{F}}(\theta)\|^2 \frac{\abs{{\bf a}^H (\theta,\phi) \cdot {\bf{i}}}^2}{{\bf{i}}^H \cdot \Re\{{\bf{Z}}\} \cdot {\bf{i}}}\label{G1},
	\end{align}
	is the array gain \cite{KD}, where 
	\begin{equation}
		{\bf a}(\theta,\phi) = \begin{bmatrix} 
			1 \\	
			e^{-jkd_2\sin\theta\cos\phi} \\
			\vdots \\
			e^{-jkd_N\sin\theta\cos\phi} \\	
		\end{bmatrix} \in \mathbb{C} ^{N \times 1} \label{a},
	\end{equation}
	is the far-field array response vector.
	The reflection (mismatch) efficiency for each dipole is:
	\begin{equation}
		\begin{aligned}	
			e_{r,n} = 1 - \abs{\Gamma_{n,c}}^2, 
		\end{aligned}\label{ern}
	\end{equation}
	where
	\begin{equation}
		\begin{aligned}
			\Gamma_{n,c} = \frac{1}{i_n}\displaystyle\sum_{m=1,m \neq n} ^{N} \Gamma_{nm} i_m \label{gcomb},
		\end{aligned}		
	\end{equation}
	is the combined reflection coefficient of each element, when all the elements are excited simultaneously \cite{comb}, while $\Gamma_{nm}$ are the reflection coefficient values extracted from the scattering matrix ${\bf S}$, found from (\ref{Zmatrix}), as delineated in \cite{pozar}.
	The total reflection efficiency $e_r$ can then be computed as in \cite{ihsan} and the realized gain is finally defined according to:
	\begin{equation}	
		\begin{aligned}		
			G_{re}(\theta,\phi) = e_r G(\theta,\phi) \label{Gre}.	
		\end{aligned}	
	\end{equation}
	
	\subsection{Parasitic Antenna Array Model}\label{sec:parmod}
	When the active element is located at $(0,0,0)$ and all the others are reactively loaded (see Fig.~\ref{dipoles}), the following modifications are made to the model. The impedance matrix of the array (\ref{Zmatrix}) is now:
	\begin{equation}	
		\begin{aligned}		
			{\bf{Z}} = {\bf{Z}}' + {\bf{X}} + {\bf R}_{loss}  \in \mathbb{C} ^{N \times N} \label{Zmatrixpar},	
		\end{aligned}
	\end{equation}
	where
	\begin{equation}
		{\bf X} = \begin{bmatrix}
			0 & 0 & \hdots & 0 \\
			0 & jX_{2} & & \vdots \\
			\vdots &  & \ddots & \vdots \\
			0 & \hdots & \hdots & jX_N
		\end{bmatrix} \in \mathbb{C} ^{N \times N} \label{Xmatrix},
	\end{equation}
	is the load matrix. In (\ref{v2}), the voltage of the parasitic dipoles is represented by $v_n=-jX_ni_n$ \cite{fast}, resulting in:
	\begin{equation}
		{\bf v} = {\bf Z} \cdot {\bf i} = \begin{bmatrix}
			v_1\\ 0\\ \vdots\\ 0 \end{bmatrix} \in \mathbb{C} ^{N \times 1} \label{v3}.
	\end{equation}
	
	The parasitic current coefficients can be found from (\ref{v3}). With all the currents known, the realized gain can then be computed from (\ref{Gre}), where the input impedance of the active dipole can be computed via (\ref{Zact}), and the reflection efficiency is as follows \cite{balanis}:
	\begin{equation}	
		\begin{aligned}
			e_r = 1 - \abs{\Gamma}^2 = 1 - \abs{\frac{Z_{in}-Z_0}{Z_{in}+Z_0}}^2 \label{er},	
		\end{aligned}	
	\end{equation}
	where $Z_0 = 73 \Omega$ is the characteristic impedance of the transmission line (or the port impedance).
	
	\section{Array Optimization}\label{sec:optimization}
\begin{figure}[htbp]
		\centering{\includegraphics[scale=0.44]{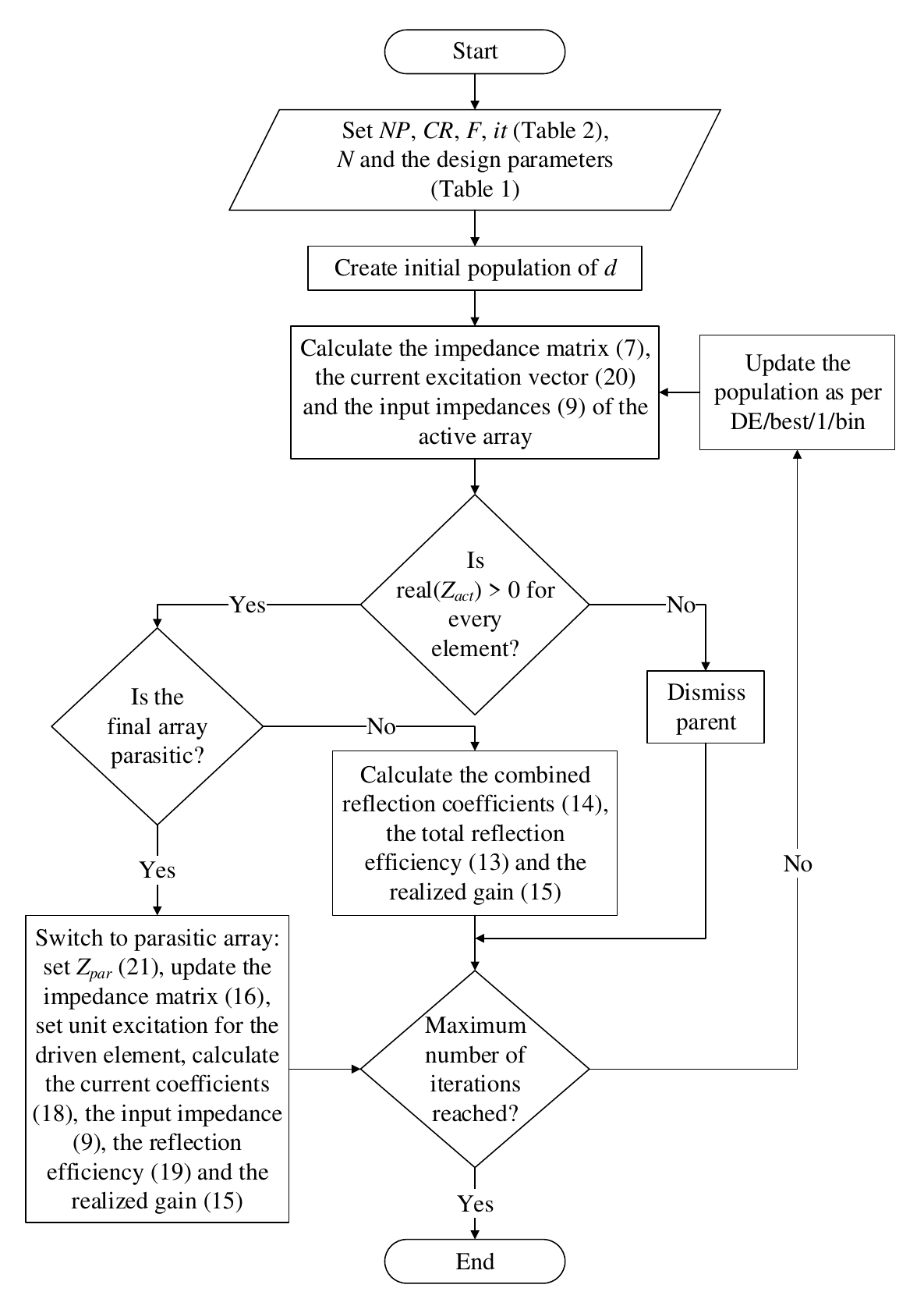}}
		\caption{Flowchart of the proposed algorithm to design either an active or a parasitic high realized gain array.}\label{flowchart}
	\end{figure}

	In this section, the proposed DE algorithm to design both an active and a parasitic array with high realized gain is presented. Figure~\ref{flowchart} depicts the corresponding flowchart of the optimization process.
	
	\subsection{Active Array Optimization} \label{optact}
	Firstly, to ensure both high directivity and gain, careful selection of the excitation currents in conjunction with the element positioning is crucial, so that the radiation efficiency remains high. To minimize the computational complexity, the current excitation of a fully-driven array that maximizes the end-fire gain is found from \cite{nossek2}, as follows:
	\begin{equation}
		\begin{aligned}
			{\bf i}_{opt} = \sqrt{\frac{N P_t}{{\bf a}^H (\theta,\phi) \cdot \Re({\bf{Z}})^{-1} \cdot {\bf a}(\theta,\phi)}} \Re({\bf{Z}})^{-1} \cdot {\bf a}(\theta,\phi) \label{iopt}.
		\end{aligned}		
	\end{equation}
	The length and radius of the unit dipole are selected to be constant, thus, the only parameter affecting (\ref{iopt}) is the spacing between the elements. Taking this into consideration, the optimization parameters including the population size $NP$, cross-over factor $CR$, mutation factor $F$, and total number of iterations $it$ are first defined (see Table~\ref{table:tableopt}). Subsequently, the initial population of $d$, which contains the element positions for $NP$ arrays, is randomly generated. The lower and upper bounds for the distance between the dipoles, $d_{ie}$, are set to $0.05\lambda$ and $0.5\lambda$, respectively. Then, the impedance matrix (\ref{Zmatrix}), excitation vector (\ref{iopt}) and input impedances (\ref{Zact}) are calculated for every parent. If any of the input impedances has a negative real part, the parent is dismissed, as it does not correspond to a practical solution. For the remaining parents, the realized gain (\ref{Gre}) is calculated according to Section \ref{sec:model}, by finding the combined reflection coefficients (\ref{gcomb}) and the total reflection efficiency (\ref{ern}). Following that, the iteration begins. In each one, the ”DE/best/1/bin” algorithm modifies the population, while applying the same condition as before and the algorithm progressively approaches the maximum achievable realized gain. The iteration process ends when the predefined total number of iterations has been reached.
	
	\begin{table}[htbp]
		\renewcommand{\arraystretch}{1.3}
		\caption{Optimization Parameters}\label{table:tableopt}
		\centering
		\begin{tabular}{| c | c | c | c |}
			\hline
			Notation & Parameter & Value$_{active}$ & Value$_{parasitic}$ \\ 
			\hline
			$NP$ & population size & $[5N,15N]$ & $[5N,13N]$ \\
			$CR$ & cross-over factor & $0.8$ & $0.8$ \\
			$F$ & mutation factor & $0.7$ & $0.7$ \\
			$it$ & number of iterations & $[20,150]$ & $[20,100]$ \\ 
			$d_{ie}$ & inter-element distance & $[0.05\lambda,0.5\lambda]$ & $[0.05\lambda,0.5\lambda]$ \\ 
			\hline
		\end{tabular}
	\end{table}
	
	The parameters of \Cref{table:tableopt} were selected so as to ensure convergence for each optimization. For instance, when the total number of elements is two ($N=2$), $NP$ and $it$ are set to their minimum values. These values then gradually increase as the array size expands, reaching their maximum for an array of seven elements ($N=7$).
	
	The study focuses on arrays with up to seven elements to balance practical applicability with performance evaluation for both active and parasitic arrays. This range effectively demonstrates the model’s capability to accurately predict and optimize the antenna configurations. In practical scenarios, both active and parasitic arrays are typically designed with a limited number of elements to meet specific antenna performance metrics such as directivity or gain, while also considering size, weight, power consumption, and cost. By restricting our investigation to seven elements, we achieve robust results while keeping the size, weight, and power consumption manageable, along with maintaining computational complexity and experimental feasibility within practical limits.
	
	The optimal active array design parameters were obtained from the algorithm described above using MATLAB and are presented in Table~\ref{table:bestactive}. The maximum inter-element distance starts at $0.27\lambda$ for $N=2$ and reaches $0.50\lambda$ for $N=5$ and above. The current amplitudes were normalized to reduce the number of utilized amplitude attenuator (AA) elements. However, except for the two-element array, $N-1$ AAs are still required, and each array requires $N$ phase shifter (PS) elements.
	
	\begin{table}[htbp]
		\renewcommand{\arraystretch}{1.3}
		\caption{Optimized Parameters for the Active Array}\label{table:bestactive}
		\centering
		\begin{tabular}{| c | c | c | c |}
			\hline
			$N$ & $n$ & $d_n/\lambda$ & ${\bf i}_{excited}$ [A]\\ 
			\hline
			$2$ & $1$ & $0$ & $1.0\cdot e^{j24.46^\circ}$ \\
			& $2$ & $0.27$ & $1.0\cdot e^{-j122.26^\circ}$ \\ 
			\hline
			& $1$ & $0$ & $0.78\cdot e^{j27.34^\circ}$ \\
			$3$ & $2$ & $0.41$ & $0.96\cdot e^{-j132.09^\circ}$ \\
			& $3$ & $0.70$ & $1.0\cdot e^{j69.78^\circ}$ \\ 
			\hline
			& $1$ & $0$ & $0.73\cdot e^{j33.27^\circ}$ \\
			$4$ & $2$ & $0.38$ & $0.69\cdot e^{-j126.44^\circ}$ \\
			& $3$ & $0.86$ & $0.92\cdot e^{j63.27^\circ}$ \\
			& $4$ & $1.14$ & $1.0\cdot e^{-j98.19^\circ}$ \\ 
			\hline
			& $1$ & $0$ & $0.68\cdot e^{j40.19^\circ}$ \\
			& $2$ & $0.37$ & $0.63\cdot e^{-j121.87^\circ}$ \\
			$5$ & $3$ & $0.83$ & $0.60\cdot e^{j67.87^\circ}$ \\
			& $4$ & $1.33$ & $0.91\cdot e^{-j102.73^\circ}$ \\
			& $5$ & $1.58$ & $1.0\cdot e^{j92.60^\circ}$ \\ 
			\hline
			& $1$ & $0$ & $0.65\cdot e^{j47.11^\circ}$ \\
			& $2$ & $0.36$ & $0.59\cdot e^{-j116.79^\circ}$ \\
			$6$ & $3$ & $0.82$ & $0.53\cdot e^{j71.68^\circ}$ \\
			& $4$ & $1.29$ & $0.55\cdot e^{-j99.30^\circ}$ \\
			& $5$ & $1.79$ & $0.92\cdot e^{j89.53^\circ}$ \\
			& $6$ & $2.02$ & $1.0\cdot e^{-j77.59^\circ}$ \\
			\hline
			& $1$ & $0$ & $0.63\cdot e^{j53.45^\circ}$ \\
			& $2$ & $0.35$ & $0.57\cdot e^{-j111.82^\circ}$ \\
			& $3$ & $0.82$ & $0.50\cdot e^{j75.52^\circ}$ \\
			$7$ & $4$ & $1.28$ & $0.49\cdot e^{-j96.27^\circ}$ \\
			& $5$ & $1.76$ & $0.52\cdot e^{j91.38^\circ}$ \\
			& $6$ & $2.26$ & $0.93\cdot e^{-j80.01^\circ}$ \\
			& $7$ & $2.48$ & $1.0\cdot e^{j111.05^\circ}$ \\
			\hline
		\end{tabular}
	\end{table}
	
	\subsection{Parasitic Array Optimization} \label{optpar}
	In order to design the parasitic array, the same steps as previously described are repeated, until any parent from the initial population with at least one active impedance exhibiting a negative real part is dismissed. Then, keeping the same positioning, all but one of the dipoles in the active array are converted to passive elements, according to \cite{activetoload}. This conversion method aims to achieve a gain comparable to the initial array while reducing the number of excitation sources. The reactive load for the $n^{th}$ element is determined based on the active element impedance, using:
	\begin{equation}
		\begin{aligned}
			Z_{par,n} = X_n = -j\Im\{Z_{act,n}\} \label{loadfromact},
		\end{aligned}		
	\end{equation}
	in order to achieve the optimal induced current magnitude and phase. To suppress the ohmic losses, the antenna is loaded only with the imaginary part of the calculated input impedance. Then, the impedance matrix of each parent is updated to (\ref{Zmatrixpar}). Given the presence of only one driven dipole, its excitation current is set to unity, eliminating the need for AAs and PSs. Subsequently, the excitation coefficients (\ref{v3}), input impedance (\ref{Zact}), reflection efficiency (\ref{er}), and, finally, the realized gain (\ref{Gre}) are found for every parent. At this stage, the iterative process begins, where the "DE/best/1/bin" algorithm updates the population over the predefined number of iterations. The process ultimately identifies the parent with the highest realized gain as the optimal solution. Similar to the active optimization, the lower parametric values of Table~\ref{table:tableopt} are reserved for the smaller array sizes and the higher ones for the larger ones.
	
	Table~\ref{table:bestparasitic} contains the parameters obtained from the optimization process. The maximum inter-element spacing ranges from $0.21\lambda$ for $N=2$ to $0.45\lambda$ for $N=7$. The optimal load is inductive for $N=2$ and capacitive for every other case. In addition, as $N$ increases, the maximum absolute and the average values of the loads rise.
	
	\begin{table}[htbp]
		\renewcommand{\arraystretch}{1.3}
		\caption{Optimized Parameters for the Parasitic Array}\label{table:bestparasitic}
		\centering
		\begin{tabular}{| c | c | c | c | c |}
			\hline
			$N$ & $n$ & $d_n/\lambda$ & ${\bf i}_{excited}$ [A] & $X_{n}$ [$\Omega$]\\ 
			\hline
			$2$ & $1$ & $0$ & $-$ & $4.07$ \\
			& $2$ & $0.21$ & $1.0$ & $-$ \\
			\hline
			& $1$ & $0$ & $-$ & $-4.18$ \\
			$3$ & $2$ & $0.36$ & $1.0$ & $-$ \\
			& $3$ & $0.57$ & $-$ & $-36.50$ \\ 
			\hline
			& $1$ & $0$ & $-$ & $-5.21$ \\
			$4$ & $2$ & $0.40$ & $1.0$ & $-$ \\
			& $3$ & $0.67$ & $-$ & $-53.31$ \\
			& $4$ & $0.98$ & $-$ & $-48.11$ \\ 
			\hline
			& $1$ & $0$ & $-$& $-4.33$  \\
			& $2$ & $0.41$ & $1.0$ & $-$ \\
			$5$ & $3$ & $0.71$ & $-$ & $-51.95$ \\
			& $4$ & $1.02$ & $-$ & $-59.67$ \\
			& $5$ & $1.43$ & $-$ & $-66.58$ \\ 
			\hline
			& $1$ & $0$ & $-$ & $-3.17$ \\
			& $2$ & $0.42$ & $1.0$ & $-$ \\
			$6$ & $3$ & $0.74$ & $-$ & $-47.85$ \\
			& $4$ & $1.03$ & $-$ & $-57.43$ \\
			& $5$ & $1.46$ & $-$ & $-88.07$ \\
			& $6$ & $1.88$ & $-$ & $-76.54$ \\
			\hline
			& $1$ & $0$ & $-$ & $-2.13$ \\
			& $2$ & $0.42$ & $1.0$& $-$  \\
			& $3$ & $0.76$ & $-$ & $-44.52$ \\
			$7$ & $4$ & $1.04$ & $-$ & $-56.41$ \\
			& $5$ & $1.46$ & $-$ & $-88.78$ \\
			& $6$ & $1.91$ & $-$ & $-104.11$ \\
			& $7$ & $2.34$ & $-$ & $-83.40$ \\
			\hline
		\end{tabular}
	\end{table}
	
	For each $N$, \Cref{table:rgcomp} provides an overview of the realized gain of the ULA and the optimized active and parasitic arrays, while \Cref{table:arraysize} presents their sizes for further examination. The ULA is steered to radiate toward the end-fire direction to ensure a fair comparison.
	
	\begin{table}[htbp]
		\renewcommand{\arraystretch}{1.3}
		\caption{Analytical End-Fire Realized Gain [dB]}\label{table:rgcomp}
		\centering
		\begin{tabular}{| c | c | c | c |}
			\hline
			$N$ & ULA$_{end-fire}$ & DE$_{active}$ & DE$_{parasitic}$ \\
			\hline
			$2$ & 4.29 & 5.85 & 6.21 \\
			$3$ & 5.43 & 7.37 & 8.17 \\
			$4$ & 6.19 & 8.36 & 9.88 \\
			$5$ & 6.77 & 9.13 & 11.06 \\
			$6$ & 7.24 & 9.76 & 11.98 \\
			$7$ & 7.63 & 10.29 & 12.72 \\
			\hline
		\end{tabular}
	\end{table}
	
	\begin{table}[htbp]
		\renewcommand{\arraystretch}{1.3}
		\caption{Size of the Array [$\lambda$]}\label{table:arraysize}
		\centering
		\begin{tabular}{| c | c | c | c |}
			\hline
			$N$ & ULA$_{end-fire}$ & DE$_{active}$ & DE$_{parasitic}$ \\ \hline
			$2$ & $0.5$ & $0.27$ & $0.21$ \\
			$3$ & $1.0$ & $0.70$ & $0.57$ \\ 
			$4$ & $1.5$ & $1.14$ & $0.98$ \\ 
			$5$ & $2.0$ & $1.58$ & $1.43$ \\
			$6$ & $2.5$ & $2.02$ & $1.88$ \\ 
			$7$ & $3.0$ & $2.48$ & $2.34$ \\ \hline
		\end{tabular}
	\end{table}
	
	\Cref{table:rgcomp} illustrates that as the number of elements in the array increases, the realized gain exhibits a consistent upward trajectory across all configurations. Upon closer examination, it is evident that the rate of increase diminishes as the number of elements grows, a phenomenon particularly noticeable in the ULA compared to the other arrays. Specifically, for the fully-driven ULA and optimized active array, the realized gain increase per additional element becomes less than 1 dB starting at $N=4$. For the parasitic array, this slowdown is first observed at $N=6$.
	
	A noteworthy observation is that, regardless of the array size, the parasitic array comprising only one active dipole consistently achieves the highest realized gain among all configurations. This suggests that the parasitic array is particularly efficient in harnessing the benefits of additional elements to enhance its performance and can maintain high reflection efficiency, a fact that will be validated in \Cref{sec:simulations}. Additionally, the optimized active array also demonstrates substantial gains, outperforming the fully-driven end-fire ULA at every element count.
	
	In terms of space requirements, the optimized arrays maintain a smaller size than the ULA for each value of $N$, as shown in \Cref{table:arraysize}, with the parasitic array being the most compact.
	
	Significantly, the optimized parasitic arrays do not require any AAs or PSs but only a single set of DAC, RF chain, and PA. On the other hand, the active ones utilize numerous DACs, RF chains, and AAs, depending on the size of the array, $N$ PSs, and $N$ PAs, while the end-fire ULA needs one set of DAC and RF chain, one PS and $N$ PAs, all of which consume power and increase the implementation cost and complexity.
	
	\section{Full-wave Simulations}\label{sec:simulations}
	\subsection{Comparison with Analytical Results}
	In this section, the results of the optimization for the active and parasitic arrays are validated with full-wave simulations in CST Microwave Studio using the Time Domain Solver. The simulation parameters are extracted from \Cref{tablemodel,table:bestactive,table:bestparasitic}. 
	
\begin{figure}[htbp]
		\centering{\includegraphics[scale=0.65]{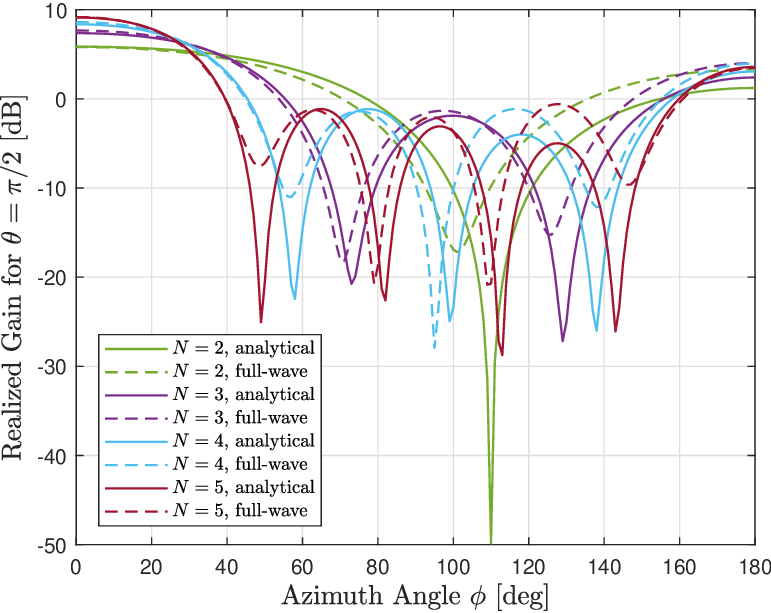}}
		\caption{The theoretical and simulated optimized realized gain of the two-, three-, four- and five-element active array.} \label{compallact}
	\end{figure}

\begin{figure}[htbp]
		\centering{\includegraphics[scale=0.65]{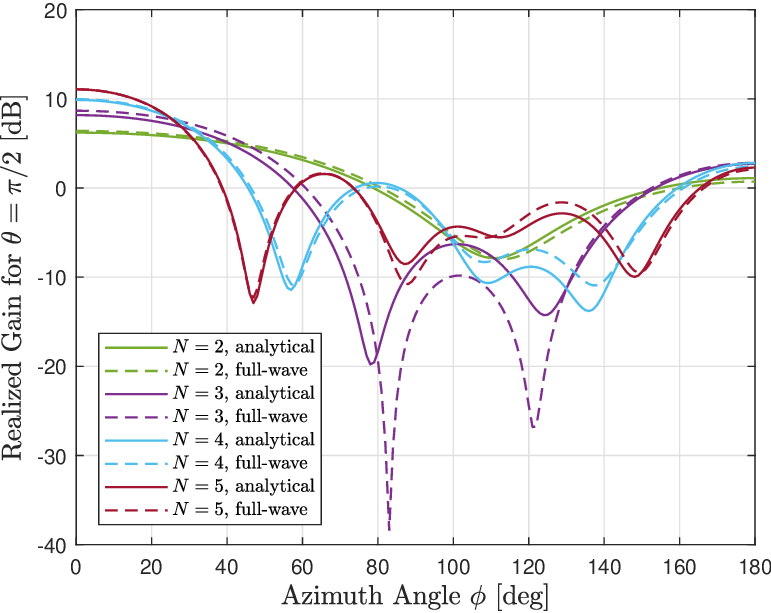}}
		\caption{The theoretical and simulated optimized realized gain of the two-, three-, four- and five-element parasitic array.} \label{compallpar}
	\end{figure}

	\Cref{table:act1,table:par1} contain the simulated realized gain and directivity values toward the end-fire direction, as well as the radiation and total efficiency of the active and parasitic arrays, respectively. The realized gain values exhibit strong agreement with the analytically obtained results in \Cref{table:rgcomp}. Notably, the two- and three-element active arrays achieve a total efficiency exceeding $80$\%, while for the parasitic arrays, it surpasses $90$\%. In the case of four- and five-element parasitic arrays, the total efficiency remains above $85$\%. The realized gain of the active and parasitic arrays is depicted in \Cref{compallact,compallpar}, respectively. To maintain clarity in these figures, the depictions for six- and seven-element arrays are omitted, but follow similar trends. The results show a strong correlation between the analytical calculations and the full-wave experiments. The relatively minor disparities can be attributed to the use of the IEP instead of the AEP, as well as the comprehensive consideration of material losses by CST, which are challenging to model analytically.
	
	\begin{table}[htbp]
		\renewcommand{\arraystretch}{1.3}
		\caption{Full-wave Simulated End-fire Directivity and Realized Gain [dB], Radiation and Total Efficiency of Active Array}\label{table:act1}
		\centering
		\begin{tabular}{| c | c | c | c | c |}
			\hline
			$N$ & Directivity [dB] & Realized Gain [dB] & $e_{cd}$ & $e_{t}$ \\
			\hline
			$2$ & 6.43 & 5.81 & 0.9958 & 0.8675 \\
			$3$ & 8.53 & 7.66 & 0.9968 & 0.8185 \\
			$4$ & 9.81 & 8.60 & 0.9959 & 0.7565 \\
			$5$ & 10.73 & 9.17 & 0.9955 & 0.6982 \\
			\hline
		\end{tabular}
	\end{table}
	
	\begin{table}[htbp]
		\renewcommand{\arraystretch}{1.3}
		\caption{Full-wave Simulated End-fire Directivity and Realized Gain [dB], Radiation and Total Efficiency of Parasitic Array}\label{table:par1}
		\centering
		\begin{tabular}{| c | c | c | c | c |}
			\hline
			$N$ & Directivity [dB] & Realized Gain [dB] & $e_{cd}$ & $e_{t}$ \\
			\hline
			$2$ & 6.66 & 6.40 & 0.9948 & 0.9401 \\
			$3$ & 8.98 & 8.66 & 0.9937 & 0.9295 \\
			$4$ & 10.52 & 9.96 & 0.9911 & 0.8779 \\
			$5$ & 11.67 & 11.06 & 0.9880 & 0.8688 \\
			\hline
		\end{tabular}
	\end{table}
	
	All evaluations were conducted on a personal computer featuring an i7-1165G7 processor, $16$ GB of RAM, and a $470$ GB SSD hard disk. The average runtime of the theoretical and simulated results as a function of $N$ is presented in \Cref{table:time}.
	
	\begin{table}[htbp]
		\renewcommand{\arraystretch}{1.3}
		\caption{Average Theoretical and Simulation Runtime [s]}\label{table:time}
		\centering
		\begin{tabular}{| c | c | c |}
			\hline
			$N$ & Avg. Theoretical Runtime [s] & Avg. Simulation Runtime [s] \\
			\hline
			$2$ & 0.132 & 42 \\
			$3$ & 0.136 & 58 \\
			$4$ & 0.146 & 73 \\
			$5$ & 0.158 & 100 \\
			\hline
		\end{tabular}
	\end{table}
	
	\Cref{table:time} demonstrates a clear contrast between the average theoretical and simulation runtimes for varying number  $N$ of elements in the antenna array. The time values presented are the averages of $10$ simulations. The theoretical runtime, computed using MATLAB, remains relatively constant, increasing only slightly from $0.132$ seconds for $N=2$ to $0.158$ seconds for $N=5$. This minimal variation suggests that the theoretical computations scale efficiently as the array size grows. In contrast, the simulation runtime, obtained using CST Studio Suite, shows a noticeable increase as the array size grows. The simulation time rises from $42$ seconds for $N=2$ to $100$ seconds for $N=5$. This aligns with typical expectations for full-wave electromagnetic simulations, where the computational complexity increases with the number of elements due to the greater number of mesh cells and interactions between array elements that need to be evaluated. In general, the simulation time is significantly larger than the corresponding theoretical time. For instance, the CST simulation for $N=2$ is approximately $318$ times longer than the theoretical runtime, while for $N=5$, it is still $633$ times longer. 
	
	It is important to note that these results reflect the runtime for a single simulation of the array for each $N$. When considering optimization processes, which involve multiple iterations of this simulation, the total runtime quickly becomes prolonged, especially for larger arrays where the full-wave simulation is most time-intensive. This highlights the critical need for efficient optimization techniques in modeling complex electromagnetic problems to manage the significant computational load of repeated electromagnetic simulations during the optimization process.
	
	\subsection{Sensitivity Analysis}
	An important aspect of the proposed optimization algorithm is its robustness to variations in the input design parameters. To assess this, the sensitivity analysis presented in \Cref{table:sensitivity} evaluates the robustness of the proposed optimization framework for the largest presented optimized design — the five-element antenna array — by examining the impact of slight variations in both the load values, $X_n$, and element position, $d_n$, on the realized gain. A $\pm5$\% scale factor is applied to each parameter, with the scale factor for the positions specifically applied to the inter-element distances rather than the absolute position of the elements, and the corresponding range of the realized gain values is recorded. 
	
	\begin{table}[htbp]
		\renewcommand{\arraystretch}{1.3}
		\begin{minipage}{\linewidth}
			\caption{Sensitivity Analysis}\label{table:sensitivity}
			\centering
			\begin{tabular}{| c | c | c |}
				\hline
				Parameter & Values* & Realized Gain [dB] \\
				\hline
				$X_1$ & [-4.55, -4.11] & 11.06 \\
				$X_3$ & [-54.55, -49.35] & 11.05 \\
				$X_4$ & [-62.66, -56.69] & [11.05, 11.07] \\
				$X_5$ & [-69.91, -63.25] & [11.05, 11.07] \\
				$d_1/\lambda$ & [-0.2, 0.2]† & 11.06 \\
				$d_2/\lambda$ & [0.39, 0.43] & [10.93, 11.07] \\
				$d_3/\lambda$ & [0.69, 0.72] & [10.97, 11.07] \\
				$d_4/\lambda$ & [1.00, 1.03] & 11.00 \\
				$d_5/\lambda$ & [1.40, 1.45] & [10.96, 11.12] \\
				\hline
			\end{tabular}
			\begin{flushleft}
				\footnotesize{* The variation factor is $\pm$5\% for this analysis.}
				\footnotesize{\newline † Since the first element is placed at the origin, $d_1$ is shifted by $\pm 5\% \cdot d_2$.}
			\end{flushleft}
		\end{minipage}
	\end{table}
	
	The realized gain values exhibit minimal sensitivity to load changes, remaining tightly clustered around $11.06$ dB, with a maximum deviation of only $0.01$ dB. This indicates that the gain is largely insensitive to small variations in the load values, which underscores the robustness of the proposed framework in handling variations in these parameters.
	
	For element positions $d_1$ to $d_5$, the realized gain shows slightly larger variations, with values ranging from a low of $10.93$ dB to a high of $11.12$ dB. This suggests that while the design's performance is somewhat more sensitive to changes in inter-element distances than in loads, the overall impact on the realized gain remains moderate and under $0.15$ dB.
	
	In conclusion, for the presented optimized design, we have demonstrated that the proposed framework is robust, maintaining stable performance even in the presence of small variations in both load values and element positions across the entire array.
	
	\section{Discussion}\label{sec:discussion}
	
	\begin{table*}[htbp]
		\renewcommand{\arraystretch}{1.3}
		\begin{minipage}{\linewidth}
			\caption{Comparison with other Parasitic Arrays}\label{table:litcomp}
			\centering
			\begin{tabular}{| c | c | c | c | c | c |}
				\hline
				References & N & Unit Element & $f$ & Directivity [dB] & Realized Gain [dB] \\
				\hline
				\cite{delav1} & 2 & Printed dipole & 850 MHz & 7.2 & - \\
				\cite{delav1} & 3 & Printed dipole & 850 MHz & 10.2 & - \\
				\cite{delav1} & 4 & Printed dipole & 844 MHz & 9.9 & -  \\
				\cite{delav2} & 4 & Printed dipole & 868 MHz & 12.5 & -18.0 \\
				\cite{umma} & 3 & EAD\textsuperscript{1} & 28 GHz & 12.31 & - \\
				\cite{DE0}* & 3 & Printed dipole & 780 MHz & 7.27 & 6.96† \\
				\cite{DE00} & 3 & Optimized\textsuperscript{2} & 980 MHz & 9.8 & - \\
				\cite{DE1} & 2 & Optimized\textsuperscript{2} & 830 / 880 MHz & 7.0 & - \\
				\cite{delav3} & 3 & Bent dipole\textsuperscript{3} & 830 MHz & 8.9 & - \\
				\cite{DE3} & 3 & Optimized\textsuperscript{2} & 920 MHz & 9.12 & 8.9 \\
				\cite{delav4} & 3 & Bent dipole & 916 MHz & - & 8.9 \\
				\cite{ziol2} & 2‡ & Optimized\textsuperscript{4} & 816 MHz & 6.75 & 6.62 \\
				This work & 2 & Wire dipole & 3.5 GHz & 6.66 & 6.40 \\
				This work & 3 & Wire dipole & 3.5 GHz & 8.98 & 8.66 \\
				This work & 4 & Wire dipole & 3.5 GHz & 10.52 & 9.96 \\
				This work & 5 & Wire dipole & 3.5 GHz & 11.67 & 11.06 \\
				\hline
			\end{tabular}
			\begin{flushleft}
				\footnotesize{* All other results were obtained through simulations, while this result is derived from theoretical analysis.}
				\footnotesize{\newline † Calculated from the information provided.}
				\footnotesize{\newline \textsuperscript{1} Egyptian Axe Dipole}
				\footnotesize{\newline \textsuperscript{2} Internally loaded wideband S-shaped monopole.}
				\footnotesize{\newline \textsuperscript{3} The fed dipole is folded.}
				\footnotesize{\newline ‡ Two resonant quadrupoles (equivalent to two pairs of resonant electric dipoles) and one magnetic dipole.}
				\footnotesize{\newline \textsuperscript{4} Bow-tie element with exceptionally high realized gain.}
			\end{flushleft}
		\end{minipage}
	\end{table*}
	
	In this section, we shift our focus to the parasitic array architecture, which has demonstrated significant advantages in terms of efficiency, compactness, and simplicity compared to the active topology previously investigated. While results for both active and parasitic arrays have been presented, the parasitic design, with only one active element, consistently achieves higher realized gain while requiring power consumption for just a single RF chain, DAC, and PA, without requiring additional PSs.
	
	The arrays developed in \Cref{sec:optimization} are compared with other parasitic arrays documented in the literature in \Cref{table:litcomp}. The table includes a variety of characteristics, such as the size of the array, the unit element used, directivity, realized gain, and efficiency. It is important to note that many references did not directly supply all the required data, and certain parameters had to be estimated based on the available results. These are noted with an asterisk (*). 
	
	For \cite{delav1}, the results with the highest achieved gain for each array size were selected: 5.8 dB, -0.7 dB, and 0.24 dB, respectively. In \cite{delav3}, the gain is 7.85 dB. The radiation efficiency was determined based on these values. The realized gain in each case is lower than the aforementioned values. For \cite{DE0}, the realized gain at $780$ MHz was found from the tables provided. For \cite{umma,DE00,DE1}, the cells corresponding to the realized gain values are left blank as these values were not provided and could not be calculated.
	
	\Cref{table:litcomp} indicates that our proposed architecture surpasses the realized gain performance at each corresponding number of elements $N$ compared to \cite{delav1, delav2}, where the narrow spacing between elements achieves high directivity but adversely affects both the gain and realized gain. Additionally, this study outperforms the theoretical results of \cite{DE0}, as well as the simulated results of \cite{delav3}.
	
	Conversely, \cite{DE3,delav4} and \cite{ziol2} report an end-fire realized gain that is marginally higher by 0.2 dB for the arrays of two and three elements. However, it is highlighted that \cite{DE3} utilizes a unit element that was internally loaded to adjust its characteristic modes, thereby optimizing the impedance bandwidth, while \cite{ziol2} incorporated a single-port single-element antenna that surpasses the maximum directivity calculated by both the Harrington and Kildal–Best size-based heuristic formulas, achieving high measured realized gain. Both of these works, as well as the SWE approach in \cite{delav4}, depend on electromagnetic simulations, starting from the unit cells. Overall, the aforementioned studies utilize unit elements that are more complex than the wire dipoles used in this work, in addition to relying on full-wave analyses that require extensive computational resources and intricate design steps. In contrast, our more straightforward approach avoids these complexities by using electromagnetic simulations solely to verify our theoretical results. Finally, there seem to be no arrays of four or more elements surpassing the results of this paper.
	
	
	\section{Conclusion}\label{sec:conclusion}
	In this paper, we introduced a novel approach for optimizing the end-fire realized gain in arrays comprising both active and parasitic dipoles. Utilizing a multi-objective DE algorithm, we developed an optimization framework to determine the optimal design parameters for arrays with two to seven elements. Specifically, we focused on optimizing the positioning and feed excitation for the active arrays, and the inter-element spacing and reactive loads for the parasitic ones. Through this process, we were able to achieve arrays with high end-fire gain and total efficiency. We examined the performance of both active and parasitic arrays in comparison to that of the phased ULA, providing key insights into the trade-offs between realized gain, total efficiency, and array size. The accuracy of the results was confirmed through full-wave electromagnetic simulations.
	
	Additionally, runtime data was incorporated into our analysis, emphasizing the efficiency of the proposed approach. By utilizing a theoretical framework as input for the optimization process, significant time savings were achieved compared to full-wave simulations, making our method computationally feasible for complex designs involving multiple iterations. Moreover, a comprehensive sensitivity analysis was conducted to assess the robustness of the optimization framework. The analysis demonstrated that the five-element array maintained stable realized gain performance despite small variations in load values and element positions, highlighting the practical reliability of the proposed algorithm in real-world applications.
	
	Our framework demonstrates superior realized gain performance compared to several parasitic arrays, as documented in the literature. While a few studies report marginally higher gains for two- and three-element arrays, they do so using significantly more complex unit elements and computationally intensive methods. In contrast, our approach achieves high performance with a simpler design and fewer computational resources. This suggests that our optimization framework not only offers a robust solution for optimizing the end-fire realized gain but also has the potential to be adapted for use with other unit elements, paving the way for future research and broader applications.
	
	Furthermore, the parasitic arrays developed in this study are ideal for base station antennas. Their compact design and high realized gain make them efficient for space-limited deployments. With only one active element, they lower power consumption and simplify hardware requirements by eliminating the need for phase shifters, RF chains, DACs, and PAs. This makes them a cost-effective and energy-efficient choice for modern base stations, positioning them as a promising option for next-generation implementations.

	\begin{IEEEbiographynophoto}{Rozita Konstantinou} is a third-year Ph.D. student in Electrical and Electronic Engineering at Queen's University Belfast. Their research is centered on advanced antenna design and optimization, particularly on parasitic antenna arrays. She holds an integrated Master's degree (MEng) in Electrical and Computer Engineering from the Aristotle University of Thessaloniki, Greece, where they graduated with Honors in 2022. Their current work emphasizes developing high-efficiency, compact antenna arrays for communication systems. 
		
	\end{IEEEbiographynophoto}
	
	\begin{IEEEbiography}[{\includegraphics[width=1in,height=1.25in,clip,keepaspectratio]{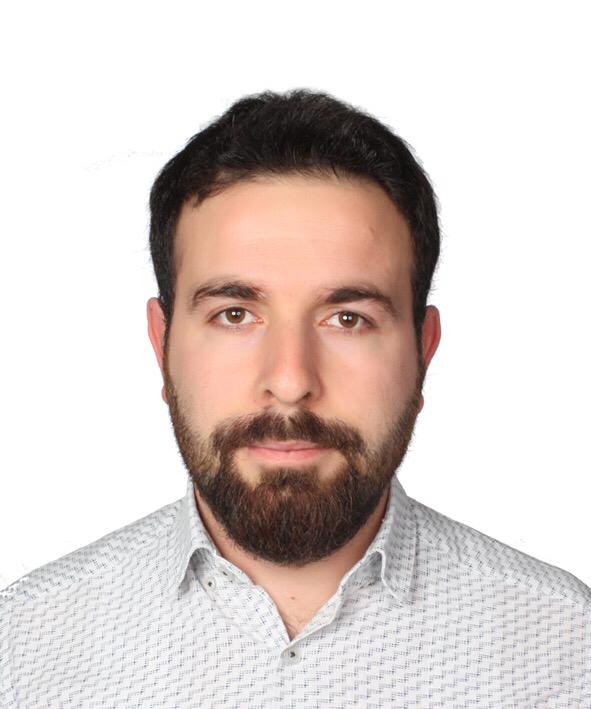}}]{Ihsan Kanbaz} (Member, IEEE) was born in Sanliurfa, Turkey, in 1991. He received the M.Sc. degree in Electrical and Electronics Engineering from Mersin University, Mersin, Turkey, in 2018 and Ph.D. degree in Electrical and Electronics Engineering from Gazi University, Ankara, Turkey.
		He is currently an Assistant Professor at Gazi University and pursuing a postdoctoral research position at Queen's University Belfast (QUB). His research interests include antenna design, test and measurement, adaptive arrays, optimization, and metaheuristic algorithms.
	\end{IEEEbiography}
	
	\begin{IEEEbiography}
		[{\includegraphics[width=1in,height=1.25in,clip,keepaspectratio]{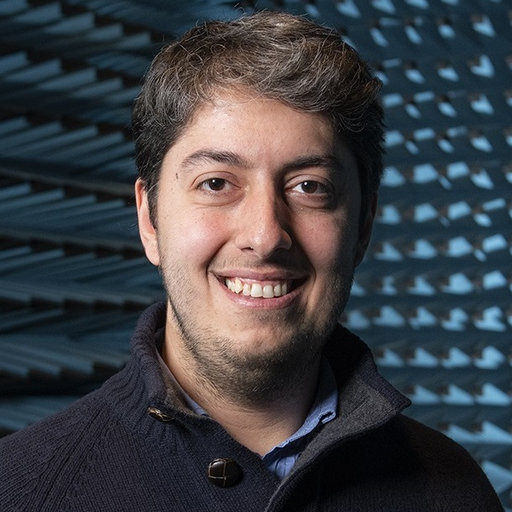}}] {Okan Yurduseven} (Senior Member, IEEE) received the Ph.D. degree in electrical engineering from Northumbria University, Newcastle upon Tyne, U.K., in 2014. From 2014 to 2018, he was a Postdoctoral Research Associate with Duke University, USA. From 2018 to 2019, he was a NASA Research Fellow with the Jet Propulsion Laboratory, California Institute of Technology, USA. He is currently a Reader (Associate Professor) with the School of Electronics, Electrical Engineering and Computer Science, Queen’s University Belfast, U.K. 
		
		Dr. Yurduseven's research has been supported with extensive funding as a Principal Investigator and a Co-Investigator (over £15 million). He has authored more than 200 peer-reviewed technical journals and conference papers. His research interests include microwave and millimeter-wave imaging, multiple-input–multiple-output (MIMO) radars, wireless power transfer, antennas, and propagation and metamaterials. 
		
		Dr. Yurduseven is a fellow of the Institution of Engineering and Technology (IET) and a member of the European Association on Antennas and Propagation (EurAAP). He was a recipient of several awards, including the Outstanding Postdoctoral Award at Duke University, in 2017, the Duke University Professional Development Award, in 2017, the NASA Postdoctoral Program Award, in 2018, the British Council—Alliance Hubert Curien Award, in 2019, the Leverhulme Trust Research Leadership Award, in 2020 (£1M), the Young Scientist Award from the Electromagnetics Academy— Photonics and Electromagnetics Research Symposium, in 2021, the Queen’s University Belfast Vice Chancellor’s Early Career Researcher Prize, in 2022, and the Outstanding Associate Editor Award from IEEE ANTENNAS AND WIRELESS PROPAGATION LETTERS, in 2023. He serves as an Associate Editor for IEEE ANTENNAS AND WIRELESS PROPAGATION LETTERS and Scientific Reports (Nature).
	\end{IEEEbiography}
	
	\begin{IEEEbiography}[{\includegraphics[width=1in,height=1.55in,clip,keepaspectratio]{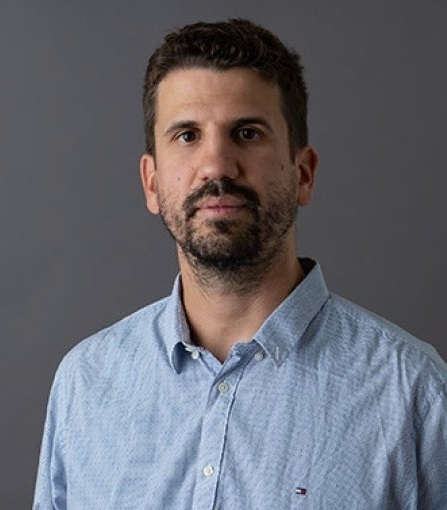}}]
		{Michail Matthaiou}(Fellow, IEEE) obtained his Ph.D. degree from the University of Edinburgh, U.K. in 2008. 
		He is currently a Professor of Communications Engineering and Signal Processing and Deputy Director of the Centre for Wireless Innovation (CWI) at Queen’s University Belfast, U.K. He has also held research/faculty positions at Munich University of Technology (TUM), Germany and Chalmers University of Technology, Sweden.
		His research interests span signal processing for wireless communications, beyond massive MIMO, reflecting intelligent surfaces, mm-wave/THz systems and AI-empowered communications.
		
		Dr. Matthaiou and his coauthors received the IEEE Communications Society (ComSoc) Leonard G. Abraham Prize in 2017. He currently holds the ERC Consolidator Grant BEATRICE (2021-2026) focused on the interface between information and electromagnetic theories. To date, he has received the prestigious 2023 Argo Network Innovation Award, the 2019 EURASIP Early Career Award and the 2018/2019 Royal Academy of Engineering/The Leverhulme Trust Senior Research Fellowship. His team was also the Grand Winner of the 2019 Mobile World Congress Challenge. He was the recipient of the 2011 IEEE ComSoc Best Young Researcher Award for the Europe, Middle East and Africa Region and a co-recipient of the 2006 IEEE Communications Chapter Project Prize for the best M.Sc. dissertation in the area of communications. He has co-authored papers that received best paper awards at the 2018 IEEE WCSP and 2014 IEEE ICC. In 2014, he received the Research Fund for International Young Scientists from the National Natural Science Foundation of China. He is currently the Editor-in-Chief of Elsevier Physical Communication, a Senior Editor for \textsc{IEEE Wireless Communications Letters} and \textsc{IEEE Signal Processing Magazine}, and an Area Editor for \textsc{IEEE Transactions on Communications}. He is an IEEE and AAIA Fellow.
	\end{IEEEbiography}
	
	\EOD
	
\end{document}